\documentclass[a4paper,10pt]{article}
\usepackage{stmaryrd}
\usepackage{amsfonts}
\usepackage{bbm}
\usepackage{amscd}
\usepackage{mathrsfs}
\usepackage{latexsym,amssymb,amsmath,amscd,amscd,amsthm,amsxtra,xypic}
\usepackage[dvips]{graphicx}
\usepackage[utf8]{inputenc}
\usepackage[T1]{fontenc}
\usepackage{lmodern}
\usepackage{amssymb}
\usepackage[all]{xy}
\usepackage{nicefrac,mathtools,enumitem}
\usepackage{microtype}

\newtheorem{thm}{Theorem}[section]
\newtheorem{lem}[thm]{Lemma}

\newtheorem{pro}[thm]{Proposition}
\newtheorem{ex}[thm]{Example}
\newtheorem{rmk}[thm]{Remark}

\newtheorem{defi}[thm]{Definition}

\newcommand{\be }{\begin{equation}}
\newcommand{\ee }{\end{equation}}

\newcommand{\pf}{\noindent{\bf Proof.}\ }

\newcommand{\frkg}{\mathfrak g}\newcommand{\g}{\mathfrak g}
\newcommand{\frkh}{\mathfrak h}

\newcommand{\frkk}{\mathfrak k}

\newcommand{\frkm}{\mathfrak m}

\def\qed{\hfill ~\vrule height6pt width6pt depth0pt}

\newcommand{\DERbr}[1]{   [    #1   ]_{\DER}   }
\newcommand{\Derbr}[1]{   [    #1   ]_{\Der}   }
\newcommand{\brh}[1]{   [    #1   ]^{\widehat{}}   }
\newcommand{\brt}[1]{   [    #1   ]'   }

\newcommand{\im}{\mathrm{\im}}

\newcommand{\dM}{\mathrm{d}}
\newcommand{\E}{\mathrm{E}}

\newcommand{\Hom}{\mathrm{Hom}}
\newcommand{\Der}{\mathrm{Der}}
\newcommand{\DER}{\mathrm{DER}}
\newcommand{\D}{\mathrm{D}}

\newcommand{\End}{\mathrm{End}}
\newcommand{\ad}{\mathrm{ad}}

\newcommand{\sgn}{\mathrm{sgn}}
\newcommand{\Ksgn}{\mathrm{Ksgn}}

\begin{document}
\title{
{Non-abelian Extensions of Lie 2-algebras
\thanks
 {The second author is supported by NSF of China (11026046,11101179), Doctoral Fund. of MEC (20100061120096) and "the Fundamental
Research Funds for the Central Universities" (200903294).  The third
author is supported by NSF of China (10971071).}}}
\author{Shaohan Chen  \\
School of Science, South China University of Technology,\\
Guangzhou 510641, Guangdong, China
\\\vspace{3mm}
email: cshjiayou@126.com\\
Yunhe Sheng  \\
Department of Mathematics, Jilin University,\\
 Changchun 130012, Jilin, China
\\\vspace{3mm}
email: shengyh@jlu.edu.cn\\
Zhujun Zheng\\
School of Science, South China University of Technology,\\
Guangzhou 510641, Guangdong, China\\
email: zhengzj@scut.edu.cn}

\date{}
\footnotetext{{\it{Keyword}:   derivations of Lie 2-algebras,
derivation Lie 3-algebra, non-abelian extensions}}
\footnotetext{{\it{MSC}}: 17B99, 53D17.} \maketitle

\begin{abstract} In this paper, we introduce the notion of derivations of Lie 2-algebras and construct the associated derivation Lie 3-algebra.
We prove that isomorphism classes of non-abelian extensions of Lie
2-algebras are classified by equivalence classes of morphisms from a
Lie 2-algebra to a derivation Lie 3-algebra.
\end{abstract}

\section{Introduction}
Eilenberg and Maclane \cite{Eilenberg} developed a theory of
non-abelian extensions of abstract groups in the 1940s, leading to
the low dimensional non-abelian group cohomology. Then there are a
lot of analogous results for Lie algebras
\cite{AMR,Nonabeliancohomology,Hochschild,Shukla}. Nonabelian
extensions of Lie algebras can be described by some linear maps
regarded as derivations of Lie algebras.  This result was
generalized to the case of super Lie algebras in \cite{AMR2}, and to
the case of Lie algebroids in \cite{brahic,Mackenzie,sheng3}.

Lie 2-algebras are the categorification of Lie algebras \cite{Baez}.
In a Lie 2-algebra, the Jacobi identity is replaced by a natural
isomorphism, which satisfies its own coherence law, called the
Jacobiator identity. The 2-category  of Lie 2-algebras is equivalent
to the 2-category of 2-term $L_\infty$-algebras, so people also view
a 2-term $L_\infty$-algebra as a Lie 2-algebra. Associated with any
Lie algebra $\frkk$,
$\frkk\stackrel{\ad}{\longrightarrow}\Der(\frkk)$ is a strict Lie
2-algebra, where $\Der(\frkk)$ is the Lie algebra of the derivations
of $\frkk$. Any non-abelian extension of a Lie algebra $\frkm$ by
$\frkk$ is described by a morphism from $\frkm$ (a trivial Lie
2-algebra) to the Lie 2-algebra
$\frkk\stackrel{\ad}{\longrightarrow}\Der(\frkk)$. Semidirect
product Lie 2-algebras and the integration of string type Lie
2-algebras were studied in \cite{sheng1}.

In this paper, we study the non-abelian extensions of Lie
2-algebras. To do that, first we develop the theory of derivations
of Lie 2-algebras. In general, for an $L_\infty$-algebra $L$, degree
$p$ derivations of $L$ is defined using coderivations of the
coalgebra
 $\wedge s(L)$ \cite{stevenson}. Concentrate on the case of Lie 2-algebras, by truncation, we construct a strict Lie 2-algebras $\Der(\g)$ associated with derivations, which
 plays important role when we consider nonabelian extensions of Lie 2-algebras.
 Motivated by the nonabelian extension theory of Lie algebras, we construct the associated strict Lie 3-algebra $\DER(\g)$, which we call the
derivation Lie 3-algebra. Any non-abelian extension of a Lie
2-algebra $\frkg$ by a Lie 2-algebra $\frkh$ gives rise to a
morphism from $\g$ to the derivation Lie 3-algebra $\DER(\frkh)$.
Furthermore, the isomorphism classes of extensions are classified by
the equivalence classes of such morphisms.

The paper is organized as follows. In Section 2, we recall some
basic definitions regarding Lie 2-algebras and strict Lie
3-algebras. In Section 3, we give the definition of derivations of
degree $0$ of Lie 2-algebras using explicit formulas. Then by
truncation, we obtain the strict Lie 2-algebra $\Der(\g)$ associated
with derivations. At last, we construct the associated strict Lie
3-algebra $\DER(\g)$, which we call the derivation Lie 3-algebra. In
Section 4, we prove that by choosing a splitting, any non-abelian
extension of the Lie 2-algebra $\g$ by $\frkh$ gives rise to a
morphism from $\frkg$ to the derivation Lie 3-algebra $\DER(\frkh)$
and different splittings give rise to equivalent morphisms.
Moreover, there is a one-to-one correspondence between the
isomorphism classes of non-abelian extensions and the equivalence
classes of morphisms.

\section{Preliminaries}
In this section, we recall some basic concepts and facts about Lie
2-algebras and strict Lie 3-algebras, and see
\cite{Baez,Lada1,Lada2} for more details. An
\emph{$L_\infty$-algebra} is a graded  vector space $L=L_0\oplus
L_1\oplus\cdots$ equipped with a system $\{l_k|~1\leq k<\infty\}$ of
linear maps $l_k:\wedge^kL\longrightarrow L$ of degree
$\deg(l_k)=k-2$, where the exterior powers are interpreted in the
graded sense and the following relation with Koszul sign ``Ksg'' is
satisfied for all $n\geq0$:
\begin{equation}
\sum_{i+j=n+1}(-1)^{i(j-1)}\sum_{\sigma}\sgn(\sigma)\Ksgn(\sigma)l_j(l_i(x_{\sigma(1)},\cdots,x_{\sigma(i)}),x_{\sigma(i+1)},\cdots,x_{\sigma(n)})=0,
\end{equation}
where the summation is taken over all $(i,n-i)$-unshuffles with
$i\geq1$. A Lie 2-algebra is a 2-term $L_\infty$-algebra. More
precisely, we have
\begin{defi} {\rm\cite{Baez}}
A Lie 2-algebra $L$ is 2-term complex of vector spaces $L: L_{1}
\stackrel{\dM}{\longrightarrow} L_{0}$  with linear maps
$\{l_k:\wedge^kL\longrightarrow L,k=2,3\}$ of degree $\deg(l_k)=k-2$
 satisfying the following equalities
\begin{itemize}
\item[$\bullet$] $\dM l_2(x,a)=l_2(x,\dM a)$,
\item[$\bullet$] $l_2(\dM a,b)=l_2(a,\dM b)$,
\item[$\bullet$] $l_2(x,l_2(y,z))+l_2(y,l_2(z,x))+l_2(z,l_2(x,y))=\dM
l_3(x,y,z)$,
\item[$\bullet$] $l_2(x,l_2(y,a))+l_2(y,l_2(a,x))+l_2(a,l_2(x,y))=l_3(x,y,\dM a)$,
\item[$\bullet$] $l_3(l_2(x,y),z,t)+c.p.=l_2(l_3(x,y,z),t)+c.p.$,
\end{itemize}
for any $x,y,z,t\in L_0,~a,b\in L_1.$ If $l_3=0$, $L$ is called a
strict Lie 2-algebra.
\end{defi}

Sometimes we use $[\cdot,\cdot]_L$ instead of $l_2$ and we denote a
Lie 2-algebra by $(L,\dM,l_2,l_3)$.

Let $ \mathbb V:V_1\stackrel{\dM}{\longrightarrow} V_0$ be a 2-term
complex of vector spaces, and we can form a new 2-term complex of
vector spaces $\End(\mathbb V):\End^1(\mathbb
V)\stackrel{\delta}{\longrightarrow} \End^0_\dM(\mathbb V)$ by
defining $\delta(D)=\dM\circ D+D\circ\dM$ for any
$D\in\End^1(\mathbb V)$, where $\End^1(\mathbb V)=\End(V_0,V_1)$ and
$$\End^0_\dM(\mathbb V)=\{X=(X_0,X_1)\in \End(V_0,V_0)\oplus \End(V_1,V_1)|~X_0\circ \dM=\dM\circ X_1\}.$$
Define  $l_2:\wedge^2 \End(\mathbb V)\longrightarrow \End(\mathbb
V)$ by setting:
\begin{equation}
\left\{\begin{array}{l}l_2(X,Y)=[X,Y]_C,\\
l_2(X,D)=[X,D]_C,\\
l_2(D,D')=0,\end{array}\right.\nonumber
\end{equation}
where $[\cdot,\cdot]_C$ is the graded commutator, for any $X,Y\in
\End^0_\dM(\mathbb V)$ and $D,D'\in \End^1(\mathbb V).$

\begin{thm}\label{thm:End(V)} {\rm\cite{Lada2,sheng2}} With the above notations,
$(\End(\mathbb V),\delta,l_2)$ is a strict Lie 2-algebra.
\end{thm}
\begin{defi}
Let $(L,\dM,l_2,l_3)$ and $(L',\dM',l_2',l_3')$ be Lie 2-algebras. A
Lie 2-algebra morphism $f:L\rightarrow L'$ consists of:
\begin{itemize}
\item[$\bullet$] two linear maps $f_{0}:L_{0}\rightarrow L_{0}'$ and $f_{1}:L_{1}\rightarrow L_{1}',$
\item[$\bullet$] one skew-symmetric bilinear map $f_{2}: L_{0}\times L_0\rightarrow L_{1}'$,
\end{itemize}
such that the following equalities hold for all $ x,y,z\in L_{0},
a\in L_{1},$
\begin{itemize}
\item[$\bullet$] $\dM'\circ f_1=f_0\circ\dM$,
\item[$\bullet$] $f_{0}l_{2}(x,y)-l_{2}'(f_{0}(x),f_{0}(y))=\dM'f_{2}(x,y),$
\item[$\bullet$] $f_{1}l_{2}(x,a)-l_{2}'(f_{0}(x),f_{1}(a))=f_{2}(x,\dM a)$,
\item[$\bullet$]
$l_2'(f_0(x),f_2(y,z))+c.p.+l_3'(f_0(x),f_0(y),f_0(z))=f_2(l_2(x,y),z)+c.p.+f_1(l_3(x,y,z))$,
\end{itemize}
where $c.p.$ means cyclic permutation. If $f_2=0$, the morphism $f$
is called a strict morphism.
\end{defi}

\begin{defi}
A strict Lie 3-algebra is a graded vector space $L=L_{0}\bigoplus
L_{1}\bigoplus L_{2}$ with linear maps
$\{l_{i}:\wedge^{i}L\rightarrow L, i=1,2\}$ of degree
$\deg(l_i)=i-2$, satisfying the following equalities for any
$x,y,z\in L$:
\begin{itemize}
\item[\rm (a)] $l_{1}^{2}=0,$
\item[\rm (b)] $l_{1}l_{2}(x,y)=l_{2}(l_{1}(x),y)+(-1)^{|x|}l_{2}(x,l_{1}(y))$,
\item[\rm (c)] $(-1)^{|x||z|}l_{2}(l_{2}(x,y),z)+(-1)^{|x||y|}l_{2}(l_{2}(y,z),x)+(-1)^{|y||z|}l_{2}(l_{2}(z,x),y)=0$.
\end{itemize}
\end{defi}

\begin{defi}\label{Def2}
Let $(L,\dM,l_{2},l_3)$ be a Lie 2-algebra and $(L',\dM',l_{2}')$ be
a strict Lie 3-algebra. A morphism $f$ from L to $L'$ consists of:
\begin{itemize}
\item[$\bullet$] two linear maps $f_{0}:L_{0}\longrightarrow L_{0}'$ and $f_{1}:L_{1}\longrightarrow L_{1}',$
\item[$\bullet$] two skew-symmetric bilinear maps $f_{2}^{0}:L_{0}\times L_0\longrightarrow L_{1}'$ and $f_{2}^{1}:L_{0}\times L_{1}\longrightarrow L_{2}',$
\item[$\bullet$] one skew-symmetric trilinear map $f_{3}: L_{0}\times L_0\times L_0\longrightarrow L_{2}',$
\end{itemize}
such that for all $ x,y,z,t\in L_{0},~a,b\in L_{1},$ we have
\begin{eqnarray*}
\dM'\circ f_{1}&=&f_{0}\circ \dM,\\
f_{0}l_{2}(x,y)-l_{2}'(f_{0}(x),f_{0}(y))&=&\dM'f_{2}^{0}(x,y),\\
f_{1}l_{2}(x,a)-l_{2}'(f_{0}(x),f_{1}(a))&=&f_{2}^{0}(x,\dM(a))+\dM'f_{2}^{1}(x,a),\\
l_{2}'(f_{1}(a),f_{1}(b))&=&f_{2}^{1}(a,\dM(b))-f_{2}^{1}(\dM(a),b),\\
f_{2}^{0}(l_{2}(x,y),z)+c.p.+f_{1}(l_{3}(x,y,z))&=&l_{2}'(f_{0}(x),f_{2}^{0}(y,z))+c.p.+\dM'f_{3}(x,y,z),\\
f_{2}^{1}(l_{2}(x,y),a)+c.p.+f_{3}(x,y,\dM
a)&=&l_{2}'(f_{0}(x),f_{2}^{1}(y,a))+l_{2}'(f_{0}(y),f_{2}^{1}(a,x))
-l_{2}'(f_{1}(a),f_{2}^{0}(x,y)),\end{eqnarray*} and
\begin{eqnarray*}
f_2^1(x,l_3(y,z,t))+l_2'(f_0(x),f_3(y,z,t))+c.p.=f_3(l_2(x,y),z,t)+c.p.+\big(l_2'(f_2^0(x,y),f_2^0(z,t))+c.p.\big).
\end{eqnarray*}
\end{defi}

\section{Derivations of Lie 2-algebras}

For a graded vector space $L$, there is a natural coalgebra
structure on $\wedge s(L)$, where $s(L)$ is the graded vector space
shifted by $1$. Another equivalent definition of an $L_\infty$
structure on $L$ is a coderivation $\partial$ of degree $-1$
satisfying $\partial^2=0$ on the coalgebra $\wedge s(L)$. See
\cite{Malte,Lada2} for more details.

\begin{defi}{\rm\cite{stevenson}}\label{defi:derivation-co}
A derivation of degree $p\geq1$ of an $L_\infty$-algebra $L$ is a
coderivation $f\in Coder^p(\wedge s(L))$ of degree $p$ of the
coalgebra $\wedge s(L)$. A derivation of degree $0$ of an
$L_\infty$-algebra $L$ is a coderivation of degree $0$ of the
coalgebra $\wedge s(L)$, which is commutative with $\partial$.
\end{defi}
Denote by $\overline{\Der^{p\geq1}}(L)$ the set of degree $p$
derivations of $L$ and $\Der^0(L)$ the set of degree $0$ derivations
of $L$, then we have a differential graded Lie algebra \cite
{stevenson}
$$
\longrightarrow\overline{\Der^p}(L)\stackrel{[\partial,\cdot]}{\longrightarrow}\cdots\stackrel{[\partial,\cdot]}{\longrightarrow}\Der^0(L)\longrightarrow0.
$$

Concentrate on the case of Lie 2-algebras, we can give the
definition of derivations of degree $0$ of Lie 2-algebras using
explicit formulas as follows.

\begin{defi}\label{Def:derivation}
Let $(\g:\g_{1}\stackrel{\dM_\frkg}{\longrightarrow}
\g_{0},[\cdot,\cdot]_\frkg,l_3^\g)$ be a Lie 2-algebra. A derivation
of degree $0$
 of $\frkg$ consists of
\begin{itemize}
  \item[$\bullet$] an element $X\in \End_{\dM}^{0}(\g)$,
  \item[$\bullet$]a skew-symmetric bilinear map $l_{X}:\g_{0}\times\g_0\longrightarrow\g_{1},$
\end{itemize}
such that for all $x,y,z\in\g_0$ and $a\in\g_1$
\begin{eqnarray*}
(a)&&X[x,y]_\frkg-[Xx,y]_\frkg-[x,Xy]_\frkg=\dM_{\g}l_{X}(x,y),\\
(b)& &X[x,a]_\frkg-[Xx,a]_\frkg-[x,Xa]_\frkg=l_{X}(x,\dM_\g a),\\
(c)& &l_{X}(x,[y,z]_\frkg)+[x,l_{X}(y,z)]_\frkg+l_{3}^\g(Xx,y,z)+l_{3}^\g(x,Xy,z)+l_{3}^\g(x,y,Xz) \nonumber \\
&&=Xl_{3}^\g(x,y,z)+l_{X}([x,y]_\frkg,z)+l_{X}(y,[x,z]_\frkg)+[l_{X}(x,y),z]_\frkg+[y,l_{X}(x,z)]_\frkg.
\end{eqnarray*}

We denote a derivation of degree $0$ of $\g$ by $(X,l_X)$ and the
set of derivations of degree $0$ of $\g$ by $\Der^{0}(\g)$.
\end{defi}

\begin{rmk}
  In a strict case,  derivations of Lie 2-algebras can be
  realized as  normalizers of the corresponding Dirac structures
  in omni-Lie 2-algebras (see Section 4 in \cite{sheng2} for more
  details).
\end{rmk}

\begin{ex}
For any $x\in \g_0$, define $\ad_x\in \End^{0}_\dM(\g)$ by
$\ad_x(y+a)=[x,y+a]_\frkg$ for any $y\in\g_0$ and $a\in\g_1$, then
$(\ad_x,l_{\ad_x}=l_3^\g(x,\cdot,\cdot))\in \Der^{0}(\g)$, which we
call an inner derivation.

\end{ex}

For any $(X,l_X),(Y,l_Y)\in \Der^{0}(\g)$, and $x,y\in\frkg_0$, we
have
\begin{eqnarray*}
&&[X,Y]_C([x,y]_\g)-[[X,Y]_C(x),y]_\g-[x,[X,Y]_C(y)]_\g\\
&=&X(Y[x,y]_\g)-Y(X[x,y]_\g)-[X(Yx)-Y(Xx),y]_\g-[x,X(Yy)-Y(Xx)]_\g\\
&=&X\big([Yx,y]_\g+[x,Yy]_\g+\dM_{\g}l_{Y}(x,y)\big)-Y\big([Xx,y]_\g+[x,Xy]_\g+\dM_{\g}l_{X}(x,y)\big)\\
&&-[X(Yx)-Y(Xx),y]_\g-[x,X(Yy)-Y(Xy)]_\g\\
&=&[X(Yx),y]_\g+[Yx,Xy]_\g+\dM_{\g}l_{X}(Yx,y)+[Xx,Yy]_\g+[x,X(Yy)]_\g+\dM_{\g}l_{X}(x,Yy)\\
&&-[Y(Xx),y]_\g-[Xx,Yy]_\g-\dM_{\g}l_{Y}(X_{0}x,y)-[Yx,Xy]_\g-[x,Y(Xy)]_\g-\dM_{\g}l_{Y}(x,Xy)\\
&&+X\dM_{\g}l_{Y}(x,y)-Y\dM_{\g}l_{X}(x,y)-[X(Yx),y]_\g
+[Y(Xx),y]_\g-[x,X(Yy)]_\g+[x,Y(Xy)]_\g\\
&=&\dM_{\g}\Big( l_{X}(Yx,y)+l_{X}(x,Yy)- l_{Y}(Xx,y)- l_{Y}(x,Xy)+
Xl_{Y}(x,y)-Yl_{X}(x,y)\Big).
\end{eqnarray*}
It is straightforward to see that
\begin{equation}\label{eq:lxy}
l_{[X,Y]_C}(x,y)\triangleq
l_{X}(Yx,y)+l_{X}(x,Yy)-l_{Y}(Xx,y)-l_{Y}(x,Xy)+Xl_{Y}(x,y)-Yl_{X}(x,y)
\end{equation}
satisfies Condition (c) in Definition \ref{Def:derivation}. Thus,
there is a well-defined bilinear skew-symmetric map
$[\cdot,\cdot]_\Der:\wedge^2\Der^0(\frkg)\longrightarrow\Der^0(\frkg):$
\begin{equation}\label{eqn:brDer1}
  [(X,l_X),(Y,l_Y)]_\Der\triangleq([X,Y]_C,l_{[X,Y]_C})
\end{equation}

  For any  $(X,l_X),(Y,l_Y), (Z,l_Z)\in \Der^{0}(\g)$, it is
  straightforward to deduce that
\begin{eqnarray*}
  l_{[X,[Y,Z]_C]_C}+l_{[Y,[Z,X]_C]_C}+l_{[Z,[X,Y]_C]_C}&=&0.
\end{eqnarray*}
Thus, we have
\begin{lem}\label{lem:derLie}
  With the above notations, $(\Der^0(\frkg),[\cdot,\cdot]_\Der)$ is
  a Lie algebra.
\end{lem}

By Definition \ref{defi:derivation-co}, the degree $1$-derivation
$\overline{\Der^1}(\g)=Coder^1(\wedge s(L))$ is given by
$$
\overline{\Der^1}(\g)=\End^1(\frkg)\oplus\End(\g_0,\wedge^2\g_0)\oplus\End(\g_1,\wedge^3\g_0).
$$
However, we find out that a smaller, thus simpler, sub-Lie 2-algebra
of the above (see Theorem \ref{thm:Lie2der}) is enough for the
application of non-abelian extensions in our setting. Thus by
truncation, we obtain a smaller Lie 2-algebra, which plays essential
role when we consider extensions of Lie 2-algebras in Section 4. To
do that, first we consider the complex
$\End^1(\frkg)\stackrel{\bar{\delta}}{\longrightarrow}\End^0(\frkg)\oplus
\Hom(\wedge^2\frkg_0,\frkg_1)$, where $\overline{\delta}$ is given
by
\begin{equation}\label{eqn:deltaba}
  \overline{\delta}(D)=(\delta(D),l_{\delta(D)}),
\end{equation}
in which $l_{\delta(D)}:\wedge^2\frkg_0\longrightarrow\frkg_1$ is
given by
\begin{equation}\label{eqn:ldeltaD}
 l_{\delta(D)}(x,y)=D[x,y]_\g-[x,D(y)]_\g-[D(x),y]_\g.
\end{equation}

\begin{pro}
  With the above notations, $\bar{\delta}(D) $ is a
  derivation, i.e. $\bar{\delta}(D)\in\Der^0(\frkg)$. Thus, we have
  a well-defined complex
\begin{equation}\label{eqn:complex}
\Der(\frkg):\Der^1(\frkg)\triangleq
\End^1(\frkg)\stackrel{\bar{\delta}}{\longrightarrow}\Der^0(\frkg).
 \end{equation}\end{pro}
\pf By \eqref{eqn:ldeltaD}, and the fact that
$\delta(D)[x,y]_\frkg=\dM_\frkg D[x,y]_\frkg$ and
$\delta(D)[x,a]_\frkg=D[x,\dM_\frkg a]_\frkg$, we have the following
two equalities obviously:
  \begin{eqnarray*}
    \delta(D)[x,y]_\g&=&[\delta(D)(x),y]_\g+[x,\delta(D)(y)]_\g+\dM_\frkg
    l_{\delta}(D)(x,y),\\
  \delta(D)[x,a]_\g&=&[\delta(D)(x),a]_\g+[x,\delta(D)(a)]_\g+ l_{\delta(D)}(x,\dM_\frkg
    a).
  \end{eqnarray*}
By straightforward computations, we can obtain Condition (c) in
Definition \ref{Def:derivation}, i.e. the following equality:
\begin{eqnarray*}
    & &l_{\delta(D)}(x,[y,z]_\frkg)+[x,l_{\delta(D)}(y,z)]_\frkg+l_{3}^\g(\delta(D)(x),y,z)+l_{3}^\g(x,\delta(D)(y),z)+l_{3}^\g(x,y,\delta(D)(z)) \nonumber \\
&&=\delta(D)l_{3}^\g(x,y,z)+l_{\delta(D)}([x,y]_\frkg,z)+l_{\delta(D)}(y,[x,z]_\frkg)+[l_{\delta(D)}(x,y),z]_\frkg+[y,l_{\delta(D)}(x,z)]_\frkg.
  \end{eqnarray*}
Thus, $\bar{\delta}(D) $ is a
  derivation. \qed\vspace{3mm}

Define a bilinear skew-symmetric map
$[\cdot,\cdot]_{\Der}:\Der^0(\frkg)\wedge\Der^1(\frkg)\longrightarrow\Der^1(\frkg)$
by:
\begin{eqnarray}\label{eqn:brDer2}
~  [(X,l_X),D]_\Der&\triangleq&[X,D]_C.
\end{eqnarray}

\begin{thm}\label{thm:Lie2der}
$(\Der(\g),\bar{\delta},[\cdot,\cdot]_\Der)$ is a strict Lie
2-algebra, when the complex $\Der(\g)$ is given by
\eqref{eqn:complex}, the differential $\bar{\delta}$ is given by
\eqref{eqn:deltaba} and the bracket $[\cdot,\cdot]_\Der$ is given by
\eqref{eqn:brDer1} and \eqref{eqn:brDer2}.
\end{thm}
\pf By Theorem \ref{thm:End(V)} and Lemma \ref{lem:derLie}, we only
need to prove that $\bar{\delta}$ is a graded derivation with
respect to the bracket operation $[\cdot,\cdot]_\Der$, i.e.
\begin{eqnarray}
\label{temp1} \bar{\delta}[(X,l_X),D]_\Der&=&[(X,l_X),\bar{\delta}(D)]_\Der,\\
 ~\label{temp2} [\bar{\delta}(D),E]_\Der&=&[D,\bar{\delta}(E)]_\Der,
\end{eqnarray}
for any $(X,l_X)\in\Der^0(\frkg)$ and $D,E\in\Der^1(\frkg)$. The
left hand side of \eqref{temp1} is equal to
\begin{eqnarray*}
  \bar{\delta}[(X,l_X),D]_\Der=
  \bar{\delta}[X,D]_C=(\delta([X,D]_C),l_{\delta([X,D]_C)}),
\end{eqnarray*}
where $l_{\delta([X,D]_C)}$ is given by
\begin{eqnarray*}
&&l_{\delta([X,D]_C)}(x,y)\\&=&[X,D]_C([x,y]_\frkg)-[[X,D]_C(x),y]_\g-[x,[X,D]_C(y)]_\g\\
&=&X\circ D[x,y]_\g-D\circ X[x,y]_\g-[x,X\circ D(y)-D\circ
X(y)]_\g-[X\circ
D(x)-D\circ X(x),y]_\g\\
&=&X([x,D(y)]_\g+[D(x),y]_\g+l_{\delta(D)}(x,y))-D[X(x),y]_\g-D[x,X(y)]_\g-D\dM_\frkg
l_X(x,y)\\
&&-[x,X\circ D(y)-D\circ X(y)]_\g-[X\circ D(x)-D\circ X(x),y]_\g\\
&=&[X(x),D(y)]_\g+l_X(x,\dM_\frkg D(y))+[D(x),X(y)]_\g+l_X(\dM_\frkg
D(x),y)+Xl_{\delta(D)}(x,y)\\
&&-D[X(x),y]_\g-D[x,X(y)]_\g-D\dM_\frkg l_X(x,y)+[x,D\circ
X(y)]_\g+[D\circ X(x),y]_\g.
\end{eqnarray*}
By \eqref{eq:lxy}, the right hand side of \eqref{temp1} is equal to
\begin{eqnarray*}
  [(X,l_X),\bar{\delta}(D)]_\Der=
 [(X,l_X),(\delta(D),l_{\delta(D)})]_\Der=([X,\delta(D)]_C,l_{[X,\delta(D)]_C}),
\end{eqnarray*}
where $l_{[X,\delta(D)]_C}$ is given by
\begin{eqnarray*}
  l_{[X,\delta(D)]_C}(x,y)&=&l_X(\dM_\frkg D(x),y)+l_X(x,\dM_\frkg
  D(y))-l_{\delta(D)}(X(x),y)-l_{\delta(D)}(x,X(y))\\
  &&+Xl_{\delta(D)}(x,y)-D
  \dM_\frkg l_X(x,y)\\
  &=&l_X(\dM_\frkg D(x),y)+l_X(x,\dM_\frkg
  D(y))-D[X(x),y]_\g+[D\circ X(x),y]_\g+[X(x),D(y)]_\g\\
  &&-D[x,X(y)]_\g+[D(x),X(y)]_\g+[x, D\circ X(y)]_\g+Xl_{\delta(D)}(x,y)-D
  \dM_\frkg l_X(x,y).
\end{eqnarray*}
Thus, we have
\begin{equation}
l_{[X,\delta(D)]_C}=l_{\delta([X,D]_C)}.
\end{equation}
Furthermore, by the fact that
$$
\delta([X,D]_C)=[X,\delta(D)]_C,
$$
we deduce that the equation \eqref{temp1} holds.

Equation \eqref{temp2} holds since we have
$[\delta(D),E]_C=[D,\delta(E)]_C$. This finishes the proof.
\qed\vspace{3mm}

In the classical case of Lie algebras, a nonabelian extension
$$
0\longrightarrow\frkk\longrightarrow\hat{\frkg}\longrightarrow\frkg\longrightarrow0
$$
can be described by a morphism from the Lie algebra $\g$ to the
strict Lie 2-algebra
$\frkk\stackrel{\ad}{\longrightarrow}\Der(\frkk)$ by choosing a
splitting. Thus, we can see that only considering the derivation Lie
algebra $\Der(\frkk)$ is not enough, we have to extend it to a Lie
2-algebra. Motivated by this, if we consider extensions of Lie
2-algebras, we have to extend the strict Lie 2-algebra $\Der(\g)$
given in Theorem \ref{thm:Lie2der} to a strict Lie 3-algebra.

Associated with the 2-term complex $\Der(\g)$, we can form a 3-term
complex of vector spaces
$$\DER(\g)\footnote{For an $L_\infty$-algebra $L$, $\Der(L)$ has been already considered by Danny Stevenson, see \cite{stevenson} for more details.}:\g_{1}\xrightarrow{\dM_\D}
\Der^{1}(\g)\oplus\g_{0}\xrightarrow {\dM_\D}\Der^{0}(\g),$$ whose
degree $0$ part $\DER^0(\g)$ is $\Der^0(\g)$, degree $1$ part
$\DER^1(\g)$ is $\Der^{1}(\g)\oplus\g_{0}$, degree $2$ part
$\DER^2(\g)$ is $\g_{1}$ and for any $a\in
\g_1,~(D,x)\in\Der^{1}(\g)\oplus\g_{0}$, $\dM_{\D}$ is given by
\begin{eqnarray*} \dM_{\D}(a)&=&(\ad-\dM_\g)(a)=(\ad_a,-\dM_\g(a)),\\
\dM_{\D}(D,x)&=&(\bar{\delta}+\ad)(D,x)=\bar{\delta}D+(\ad_x,l_{\ad_x}).\end{eqnarray*}
$\dM_\D^2=0$ follows from \begin{equation}\label{eq:deltaad}
\delta(\ad_a)=\ad_{\dM_\g a}.
\end{equation}

Define a bilinear degree $0$ bracket $
\DERbr{\cdot,\cdot}:\DER(\g)\wedge\DER(\g)\longrightarrow \DER(\g)$
by
\begin{equation}\label{2bracket}
\left\{\begin{array}{rll}\DERbr{(X,l_X),(Y,l_Y)}&=&\Derbr{(X,l_X),(Y,l_Y)},\\
\DERbr{(X,l_X),(D,x)}&=&\big(\Derbr{(X,l_X),D}+l_X(x,\cdot),X(x)\big),\\
\DERbr{(D,x),(D',x')}&=&-Dx'-D'x,\\
\DERbr{(X,l_X),a}&=&X(a),
\end{array}\right.
\end{equation}
for any  $(X,l_X),(Y,l_Y)\in \DER^{0}(\g)$, $(D,x),(D',x')\in
\DER^{1}(\mathfrak{g})$ and $a\in \DER^2(\g)$.

\begin{thm}
With the above notations,
$(\DER(\mathfrak{g}),\dM_\D,\DERbr{\cdot,\cdot})$ is a strict Lie
3-algebra, which we call the derivation Lie 3-algebra of
$\mathfrak{g}$.
\end{thm}

\pf We only need to show that $\dM_\D$ is a graded derivation with
respect to the bracket operation $\DERbr{\cdot,\cdot}$, and
$\DERbr{\cdot,\cdot}$ satisfies the graded Jacobi identity. The
condition  that $\dM_\D$ is a graded derivation is equivalent to
\begin{eqnarray}
\label{1}\dM_\D\DERbr{(X,l_X),a}&=&\DERbr{(X,l_X),\dM_\D(a)},\\
\label{2}\dM_\D\DERbr{(X,l_X),(D,x)}&=&\DERbr{(X,l_X),\dM_\D(D,x)},\\
\label{3}\dM_\D\DERbr{(D,x),(D',x')}&=&\DERbr{\dM_\D(D,x),(D',x')}
-\DERbr{(D,x),\dM_\D(D',x')},\\
 \label{33} \DERbr{\dM_\D(D,x),a}&=& \DERbr{(D,x),\dM_\D
 a}.
\end{eqnarray}
The left hand side of \eqref{1} is equal to $(\ad_{X(a)},-\dM_\frkg
X(a))$, and the right hand side is equal to
$$([X,\ad_a]_C-l_X(\dM_\frkg a,\cdot),-X(\dM_\frkg a)).$$ By
the fact that $[X,\ad_a]_C=\ad_{X(a)}+l_X(\dM_\frkg a,\cdot)$, we
obtain \eqref{1}.

The left hand side of \eqref{2} is equal to
\begin{eqnarray*}
&&\dM_\D\DERbr{(X,l_X),(D,x)}\\
&=&\dM_\D\Big(\Derbr{(X,l_X),D}+l_X(x,\cdot),X(x)\Big)\\
&=&\dM_\D\Big([X,D]_C+l_X(x,\cdot),X(x)\Big)\\
&=&\Big(\delta\big([X,D]_C+l_X(x,\cdot)\big)+\ad_{Xx},l_{\delta[X,D]_C}+l_{\delta(l_X(x,\cdot))}+l_{\ad_{X(x)}}\Big).
\end{eqnarray*}
The right hand side of \eqref{2} is equal to
\begin{eqnarray*}
&&\DERbr{(X,l_X),\dM_\D(D,x)}\\
&=&\Derbr{(X,l_X),(\delta(D),l_{\delta(D)})+(\ad_x,l_{\ad_x})}\\
&=&([X,\delta(D)+\ad_x]_C,l_{[X,\delta(D)+\ad_x]_C})\\
&=&\big([X,\delta(D)]_C+\ad_{X(x)}+\delta(l_X(x,\cdot)),l_{[X,\delta(D)]_C+\ad_{X(x)}+\delta(l_X(x,\cdot))}\big).
\end{eqnarray*}
The last equality holds since $(X,l_X)$ is a derivation. Therefore,
by the fact that $\delta$ is a graded derivation with respect to the
bracket operation $[\cdot,\cdot]_C$, we deduce that
$$
\dM_\D\DERbr{(X,l_X),(D,x)}=\DERbr{(X,l_X),\dM_\D(D,x)}.
$$

The left hand side of \eqref{3} is equal to
\begin{eqnarray*}
\dM_D(-Dx'-D'x)=(-\ad_{D(x')}-\ad_{D'(x)},\dM_\frkg(D(x'))+\dM_\frkg(D'(x))).
\end{eqnarray*}
The right hand side of \eqref{3} is equal to
\begin{eqnarray*}
&&\DERbr{\delta(D)+\ad_x,(D',x')}-\DERbr{(D,x),\delta(D')+\ad_{x'}}\\
&=&\Big([\delta(D),D']_C+[\ad_x,D']_C+l_{\delta(D)}(x',\cdot)+l_{\ad_x}(x',\cdot),\delta(D)(x')+[x,x']_\frkg\Big)\\
&&-\Big([D,\delta(D')]_C+[D,\ad_{x'}]_C-l_{\delta(D')}(x,\cdot)-l_{\ad_{x'}}(x,\cdot),-\delta(D')(x)-[x',x]_\frkg\Big)\\
&=&\Big([\ad_x,D']_C+l_{\delta(D')}(x,\cdot)-[D,\ad_{x'}]_C+l_{\delta(D)}(x',\cdot),\dM_\frkg(D(x'))+\dM_\frkg(D'(x))\Big).
\end{eqnarray*}
By \eqref{eqn:ldeltaD}, we deduce that \eqref{3} holds. It is
straightforward to deduce that \eqref{33} holds.

The bracket operation $\DERbr{\cdot,\cdot}$ satisfies graded Jacobi
identity, and it is equivalent to
\begin{eqnarray}\label{4}
\DERbr{\DERbr{(X,l_X),(Y,l_Y)},a}+c.p.&=&0,\\
\label{5} \DERbr{\DERbr{(X,l_X),(Y,l_Y)},(D,x)}+c.p.&=&0,
\end{eqnarray}
and
\begin{eqnarray}\label{6}
&&\nonumber\DERbr{\DERbr{(X,l_X),(D,x)},(D',x')}+\DERbr{\DERbr{(D,x),(D',x')},(X,l_X)}\\
&&-\DERbr{\DERbr{(D',x'),(X,l_X)},(D,x)}=0.
\end{eqnarray}
It is obvious that \eqref{4} holds. By straightforward computations,
the left hand side of \eqref{5} is equal to
\begin{eqnarray*}
 &&\DERbr{([X,Y]_C,l_{[X,Y]_C}),(D,x)}+\DERbr{([Y,D]_C+l_Y(x,\cdot),Yx),(X,l_X)}\\
&&
+\DERbr{([D,X]_C-l_X(x,\cdot),-Xx),(Y,l_Y)}\\
&=&\Big([[X,Y]_C,D]_C+l_{[X,Y]_C}(x,\cdot),[X,Y]_C(x)\Big)\\
&&+\Big([[Y,D]_C+l_Y(x,\cdot),X]_C-l_{X}(Yx,\cdot),-X(Yx)\Big)\\
&&+\Big([[D,X]_C-l_X(x,\cdot),Y]_C+l_{Y}(Xx,\cdot),Y(Xx)\Big).
\end{eqnarray*}
Since $[\cdot,\cdot]_C$ satisfies the Jacobi identity and by the
definition of $l_{[X,Y]_C}$ (see \eqref{eq:lxy}), we get \eqref{5}.
\eqref{6} can be deduced similarly.
 \qed\vspace{3mm}

\begin{defi}\label{defi:equivalent}
Let
$(\frkg,\dM_\g,[\cdot,\cdot]_\g,l_3^\g),~(\frkh,\dM_\frkh,[\cdot,\cdot]_\frkh,l_3^\frkh)$
be two Lie 2-algebras. Assume that
$f=(f_{0},f_{1},f_{2}^{0},f_{2}^{1},f_3)$ and
$f'=(f_{0}',f_{1}',{f_{2}^{0}}',{f_{2}^{1}}',f_3')$ are two
morphisms from  $\g$ to  $\DER(\frkh)$. We say that $f'$ is
equivalent to $f$ if there exist:
\begin{itemize}
\item[$\bullet$] linear maps $b_{0}:\g_{0}\longrightarrow \frkh_{0}$ and  $b_{1}:\g_{1}\longrightarrow\frkh_{1},$
\item[$\bullet$] a bilinear map $b_2:\wedge^{2}\g_{0}\longrightarrow \frkh_{1},$
\end{itemize}
such that $(b_0,b_1)$ is a chain homotopy from $(f_0',f_1')$ to
$(f_0,f_1)$:
\begin{eqnarray*}
\label{e1}f_{0}-f_{0}'&=&\dM_\D\circ b_{0},\\
\label{e2}f_{1}-f_{1}'&=&b_{0}\circ \dM_\g+\dM_\D\circ b_{1},
\end{eqnarray*}
and the following equalities hold for all $x,y,z\in
\mathfrak{g}_{0}$ and $a\in \mathfrak{g}_{1},$
\begin{eqnarray*}
\label{e4}({f_2^0}'-f_2^0)(x,y)&=&\DERbr{f_0'(x),b_0(y)}-\DERbr{f_0'(y),b_0(x)}-b_0([x,y]_\g)\nonumber\\
&&+\DERbr{\dM_\D b_0(x),b_0(y)}-\dM_\D b_2(x,y),\\
\label{e5}({f_2^1}'-f_2^1)(x,a)&=&\DERbr{f_{0}'(x),b_{1}(a)}+\DERbr{f_1'(a),b_{0}(x)}-b_{1}([x,a]_\g)\nonumber\\
&&+\DERbr{\dM_\D b_{0}(x),b_{1}(a)}+b_2(x,\dM_\mathfrak{g}a),\\
\label{e3} (f_3'-f_3)(x,y,z)&=&\DERbr{f_0'(x),b_2(y,z)}-b_2([x,y]_\g,z)+c.p.\\
\nonumber&&+\DERbr{{f_2^0}'(x,y),b_0(z)}+\DERbr{\dM_\D b_0(x),b_2(y,z)}+l_{f_0'(x)}(b_0(y),b_0(z))+c.p.\\
\nonumber&&-b_1(l_3^\g(x,y,z))+l_3^\frkh(b_0(x),b_0(y),b_0(z)).
\end{eqnarray*}
\end{defi}

\begin{rmk}
  Let
  $\frkh=(0\stackrel{0}{\longrightarrow}\frkh_0,[\cdot,\cdot]_{\frkh_0},l_3=0)$
  be the trivial Lie 2-algebra determined by a Lie algebra
  $\frkh_0$, then the Lie 3-algebra $\DER(\frkh)$ reduces to the
  well-known Lie 2-algebra
  $\frkh_0\stackrel{\ad}{\longrightarrow}{\Der(\frkh_0)}$. Two
  morphisms $f=(f_0,f_1,f_2^0)$ and $f'=(f_0',f_1',{f_2^0}')$ from $\frkg$ to $\frkh_0\stackrel{\ad}{\longrightarrow}{\Der(\frkh_0)}$ are
  equivalent if and only if there is a linear map
  $b_0:\g_0\longrightarrow\frkh_1$ such that
  \begin{eqnarray*}
    f_0-f_0'&=&\ad\circ b_0,\\
    f_1-f_1'&=&b_0\circ\dM_\g,\\
   ({
   f_2^0}'-f_2^0)(x,y)&=&f_0'(x)(b_0(y))-f_0'(y)(b_0(x))-b_0([x,y]_\g)+[b_0(x),b_0(y)]_\frkh,
  \end{eqnarray*}
  i.e. $b_0$ is a 2-morphism from $f'$ to $f$ in the sense of Baez-Crans \cite{Baez}.
\end{rmk}

\section{Non-abelian Extensions of Lie 2-algebras}
 \begin{defi}
 \begin{itemize}
 \item[\rm (i)] Let $\mathfrak{g}:\mathfrak{g}_{1}\longrightarrow\mathfrak{g}_{0}$, $\mathfrak{h}:\frkh_{1}\longrightarrow\frkh_{0}$, $\hat{\mathfrak{g}}:\hat{\mathfrak{g}}_{1}
\longrightarrow\hat{\mathfrak{g}}_{0}$ be Lie 2-algebras and
$i=(i_{1},i_{0}):\frkh\longrightarrow\hat{\mathfrak{g}},~~p=(p_{1},p_{0}):\hat{\mathfrak{g}}\longrightarrow\mathfrak{g}$
be strict morphisms. The following sequence of Lie 2-algebras is a
short exact sequence if $\mathrm{Im}(i)=\mathrm{Ker}(p)$,
$\mathrm{Ker}(i)=0$ and $\mathrm{Im}(p)=\g$.

\begin{equation}\label{eq:ext1}
\CD
  0 @>0>>  \frkh_1 @>i_1>> \hat{\g}_1 @>p_1>> \g_1 @>0>> 0 \\
  @V 0 VV @V \dM_\frkh VV @V \hat{\dM} VV @V\dM_\g VV @V0VV  \\
  0 @>0>> \frkh_{0} @>i_0>> \hat{\g}_0 @>p_0>> \g_0@>0>>0
\endCD
\end{equation}

We call $\hat{\mathfrak{g}}$  an extension of $\mathfrak{g}$ by
$\frkh$, and denote it by $\E_{\hat{\g}}.$
\item[\rm (ii)] A splitting $\sigma:\mathfrak{g}\longrightarrow\hat{\mathfrak{g}}$ of $p:\hat{\mathfrak{g}}\longrightarrow\mathfrak{g}$
consists of linear maps
$\sigma_0:\mathfrak{g}_0\longrightarrow\hat{\g}_0$ and
$\sigma_1:\mathfrak{g}_1\longrightarrow\hat{\g}_1$
 such that  $p_0\circ\sigma_0=id_{\mathfrak{g}_0}$ and  $p_1\circ\sigma_1=id_{\mathfrak{g}_1}$.
\item[\rm (iii)] We say that two extensions of Lie 2-algebras
 $\E_{\hat{\g}}:\frkh\stackrel{i}{\longrightarrow}\hat{\g}\stackrel{p}{\longrightarrow}\g$
 and $\E_{\tilde{\g}}:\frkh\stackrel{j}{\longrightarrow}\tilde{\g}\stackrel{q}{\longrightarrow}\g$ are isomorphic
 if there exists a Lie 2-algebra morphism $F:\hat{\g}\longrightarrow\tilde{\g}$  such that $F\circ i=j$, $q\circ
 F=p$ and $F_2(i(u),\alpha)=0$, for any
 $u\in\frkh_0,~\alpha\in\hat{\g}_0$.
\end{itemize}
\end{defi}

In the sequel, we will write an element $(X,l_X)\in\Der^0(\frkh)$,
by $X$ to simplify the computation. \vspace{2mm}

Given a splitting $\sigma$, we have
$\hat{\g}_0\cong\g_0\oplus\frkh_0$ and
$\hat{\g}_1\cong\g_1\oplus\frkh_1$ as vector spaces. Furthermore,
$(i_0,i_1)$ are inclusions and $(p_0,p_1)$ are projections. $\sigma$
induces linear maps:
$$
\begin{array}{rlclcrcl}
 \varphi:&\mathfrak{g}_{1}&\longrightarrow&\frkh_{0},&& \varphi(a)&\triangleq&\hat{\dM}\sigma(a)-\sigma(\dM_\g a),\\
\mu_{0}:&\mathfrak{g}_{0}&\longrightarrow& \Der^{0}(\frkh),&& \mu_{0}(x)(u+m)&\triangleq&[\sigma(x),u+m]_{\hat{\mathfrak{g}}},\\
\mu_{1}:&\mathfrak{g}_{1}&\longrightarrow& \Der^{1}(\frkh),&&\mu_{1}(a)(u)&\triangleq&[\sigma(a),u]_{\hat{\mathfrak{g}}},\\
\mu_{2}:&\wedge^2\mathfrak{g}_{0}&\longrightarrow&\Der^{1}(\frkh),&& \mu_{2}(x,y)&\triangleq&\hat{l_{3}}(\sigma(x),\sigma(y),\cdot),\\
\omega:&\wedge^2\mathfrak{g}_{0}&\longrightarrow&\frkh_{0},&& \omega(x,y)&\triangleq&\sigma[x,y]_\mathfrak{g}-[\sigma(x),\sigma(y)]_{\hat{\mathfrak{g}}},\\
\nu:&\mathfrak{g}_{0}\wedge\mathfrak{g}_{1}&\longrightarrow&\frkh_{1},&& \nu(x,a)&\triangleq&\sigma[x,a]_\mathfrak{g}-[\sigma(x),\sigma(a)]_{\hat{\mathfrak{g}}},\\
\theta:&\wedge^3\mathfrak{g}_{0}&\longrightarrow&\frkh_{1},&&
\theta(x,y,z)&\triangleq&\sigma(l_{3}^\g(x,y,z))-\hat{l_{3}}(\sigma(x),\sigma(y),
\sigma(z)),
\end{array}
$$
for any $x,y,z\in\mathfrak{g}_{0}$, $a\in\mathfrak{g}_{1}$,
$u\in\frkh_{0}$ and $m\in\frkh_{1}$.

\begin{pro} \label{prop1}
The splitting $\sigma$ induces a morphism
\begin{equation}\label{eq:morphism}
f=(f_0,f_1,f_2^0,f_2^1,f_3)=(\mu_{0},\mu_{1}-\varphi,-\mu_{2}+\omega,\nu,\theta)
\end{equation}
from the Lie 2-algebra $\mathfrak{g}$ to the derivation Lie
3-algebra $\DER(\frkh)$. Moreover, different splittings give
equivalent morphisms.
\end{pro}
\pf By computations, we have
\begin{eqnarray*}
\big(\dM_\D\circ(\mu_1-\varphi)(a)\big)(u+m)&=&(\delta+\ad)(\mu_1(a),-\varphi(a))(u+m)\\
&=&\delta(\mu_1(a))(u+m)-\ad_{\varphi(a)}(u+m)\\
&=&\dM_\frkh[\sigma(a),u]_{\hat{\g}}+[\sigma(a),\dM_\frkh
m]_{\hat{\mathfrak{g}}} -[\varphi(a),u+m]_\frkh\\
&=&[\hat{\dM}\sigma(a),u]_{\hat{\mathfrak{g}}}+[\hat{\dM}\sigma(a),m]_{\hat{\mathfrak{g}}}-[\varphi(a),u+m]_\frkh \\
&=&[\sigma(\dM_\g a)+\varphi(a),u+m]_{\hat{\mathfrak{g}}}-[\varphi(a),u+m]_\frkh \\
&=&\mu_{0}(\dM_\g a)(u+m),
\end{eqnarray*} which implies that \begin{equation}\label{eq:con1}\dM_\D\circ
f_1=f_{0}\circ\dM_\g.\end{equation}

We have the equalities
\begin{eqnarray}
\label{eqn:xyu}&&[\sigma x,[ \sigma y, u]_{\hat{\mathfrak{g}}}]_{\hat{\mathfrak{g}}}+c.p.=\hat{\dM}\hat{l_{3}}(\sigma x,\sigma y,u), \\
\label{eqn:xym}&&[\sigma x, [\sigma y,
m]_{\hat{\mathfrak{g}}}]_{\hat{\mathfrak{g}}}+c.p.=\hat{l_{3}}(\sigma
x,\sigma y,\dM_\frkh m).
\end{eqnarray}
The left hand side of (\ref{eqn:xyu}) is equal to
\begin{eqnarray*}
&&[u,\sigma[x,y]_{\g}-\omega(x,y)]_{\hat{\g}}+[\sigma(x),\mu_0(y)u]_{\hat{\g}}-[\sigma(y),\mu_0(x)u]_{\hat{\g}}\\
&=&\DERbr{\mu_0(x),\mu_0(y)}(u)-\mu_0([x,y]_\g)(u)+\ad_{\omega(x,y)}(u),
\end{eqnarray*}
and the right hand side is equal to $\dM_\frkh(\mu_{2}(x,y)u),$
which implies that
$$
\DERbr{\mu_0(x),\mu_0(y)}(u)-\mu_0([x,y]_\g)(u)=\dM_\frkh(\mu_{2}(x,y)(u))-\ad_{\omega(x,y)}u.
$$
Similarly, by \eqref{eqn:xym}, we get
$$
\DERbr{\mu_0(x),\mu_0(y)}(m)-\mu_0([x,y]_\g)(m)=\mu_{2}(x,y)(\dM_\frkh
m)-\ad_{\omega(x,y)}m.
$$
Therefore, we have
\begin{eqnarray}
  \label{eq:con2}\nonumber f_0([x,y]_\g)-\DERbr{f_0(x),f_0(y)}&=&\mu_0([x,y]_\g)-\DERbr{\mu_0(x),\mu_0(y)}\\
  &=&-\delta(\mu_{2}(x,y))+\ad_{\omega(x,y)}=\dM_\D\circ f_2^0(x,y).
\end{eqnarray}

We have the equality
\begin{eqnarray*}
&&[\sigma x,[\sigma a,
u]_{\hat{\mathfrak{g}}}]_{\hat{\mathfrak{g}}}+c.p.=\hat{l_{3}}(\sigma
x,\hat{\dM}\sigma a,u).
\end{eqnarray*}
Thus, we have
\begin{eqnarray*}
&&\DERbr{\mu_0 (x),\mu_1
(a)}(u)-\mu_1([x,a]_\g)(u)+\ad_{\nu(x,a)}u\\
&=&[\sigma x,\mu_1( a)(
u)]_{\hat{\mathfrak{g}}}+[\sigma(a),-\mu_0(x)(u)]_{\hat{\g}}+[u,\sigma[x,a]_\g-\nu(x,a)]_{\hat{\g}}\\
&=&\hat{l_{3}}(\sigma x,\sigma \dM_\g(a)+\varphi(a),u)\\
&=&\mu_2(x, \dM_\g(a))(u)+\hat{l_{3}}(\sigma x,\varphi(a),u).
\end{eqnarray*}
By the equality $\hat{\dM}[\sigma x,\sigma
a]_{\hat{\mathfrak{g}}}=[\sigma x,\hat{\dM}\sigma
a]_{\hat{\mathfrak{g}}}$, we obtain that
\begin{eqnarray}
\label{eqn:omeganu}\mu_0(x)(\varphi(a))-\varphi([x,a]_\mathfrak{g})&=&\omega(x,\dM_\mathfrak{g}a)-\dM_\frkh\nu(x,a).
\end{eqnarray}
Therefore, we have
\begin{eqnarray}
\label{eq:con3}&&f_1([x,a]_\g)-\nonumber\DERbr{f_0(x),f_1(a)}\\
\nonumber&=&\mu_1([x,a]_\g)-\varphi([x,a]_\g)-\DERbr{\mu_0(x),\mu_1(a)-\varphi(a)}\\
\nonumber&=&\mu_1([x,a]_\g)-\Derbr{\mu_0(x),\mu_1(a)}+l_{\mu_0(x)}(\varphi(a),\cdot)+\mu_0(x)(\varphi(a))-\varphi([x,a]_\g)\\
\nonumber&=&-\mu_2(x,
\dM_\g a)-\hat{l_{3}}(\sigma x,\varphi(a),\cdot)+\ad_{\nu(x,a)}+l_{\mu_0(x)}(\varphi(a),\cdot)+\omega(x,\dM_\mathfrak{g}a)-\dM_\frkh\nu(x,a)\\
&=&f_2^0(x,\dM_\g a)+\dM_\D f_2^1(x,a).
\end{eqnarray}

By the equality $[\hat{\dM}\sigma a,\sigma
b]_{\hat{\mathfrak{g}}}=[\sigma a,\hat{\dM}\sigma
b]_{\hat{\mathfrak{g}}},$ we obtain that
\begin{eqnarray}
  \label{eq:con4}\nonumber\DERbr{f_1(a),f_1(b)}&=&\DERbr{\mu_1(a)-\varphi(a),\mu_1(b)-\varphi(b)}=\mu_{1}(a)\varphi(b)+\mu_{1}(b)\varphi(a)\\
  &=&\nu(a,\dM_\g
b)-\nu(\dM_\g a,b)=f_2^1(a,\dM_\g b)-f_2^1(\dM_\g a,b).
\end{eqnarray}

By the equality
$$
[\sigma x, [\sigma y,\sigma
z]_{\hat{\mathfrak{g}}}]_{\hat{\mathfrak{g}}}+c.p.=\hat{\dM}\hat{l_{3}}(\sigma
x,\sigma y,\sigma z),
$$
we get
\begin{equation}\label{eq:mu0omega}
-\mu_{0}(x)\omega(y,z)-\omega(x,[y,z]_\mathfrak{g})+c.p.=-\dM_\frkh
\theta(x,y,z)+\varphi(l_3^\g(x,y,z)).
\end{equation}
By the Jacobiator identity:
\begin{eqnarray*}
\hat{l_{3}}([\sigma x,\sigma y]_{\hat{\g}},\sigma z,u)+c.p.=[\sigma
x,\hat{l_{3}}(\sigma y,\sigma z,u)]_{\hat{\mathfrak{g}}}+c.p.,
\end{eqnarray*}
we have
\begin{equation}\label{eqn:mu0mu2}
 [\mu_0(x),\mu_2(y,z)]_{\DER}-l_{\mu_0(x)}(\omega(y,z),\cdot)+c.p.=\mu_2([x,y]_\g,z)+c.p.+\ad_{\theta(x,y,z)}
-\mu_1l_3^\g(x,y,z).
\end{equation}
By \eqref{eq:mu0omega} and \eqref{eqn:mu0mu2}, we have
\begin{eqnarray}
\label{eq:con5}\nonumber&&\DERbr{f_0(x),f_2^0(y,z)}+c.p.+\dM_\D f_3(x,y,z)\\&=&\nonumber\DERbr{\mu_{0}(x),(-\mu_2+\omega)(y,z)}+c.p.+\dM_\D\theta(x,y,z)\\
\nonumber&=&\big(-[\mu_{0}(x),\mu_2(y,z)]_\Der+l_{\mu_{0}(x)}(\omega(y,z),\cdot)+\mu_{0}(x)\omega(y,z)+c.p.\big)+\dM_\D\theta(x,y,z)\\
\nonumber&=&\big(-\mu_2([x,y]_\g,z)+\omega([x,y]_\g,z)+c.p.\big)+\mu_1l_3^\g(x,y,z)-\varphi(l_3^\g(x,y,z))\\
&=&f_2^0([x,y]_\g,z)+c.p.+f_1l_3^\g(x,y,z).
\end{eqnarray}

By the equality
\begin{eqnarray}
\label{eqn:xya}&&[\sigma x, [\sigma y,\sigma
a]_{\hat{\mathfrak{g}}}]_{\hat{\mathfrak{g}}}+c.p.=\hat{l_{3}}(\sigma
x,\sigma y,\hat{\dM}\sigma a),
\end{eqnarray}
we have
\begin{eqnarray*}
&&[\sigma x,\sigma[y,a]_\g-\nu(y,a)]_{\hat{\g}}+[\sigma y,\sigma[a,x]_\g-\nu(a,x),]_{\hat{\g}}+[\sigma a,\sigma[x,y]_\g-\omega(x,y)]_{\hat{\g}}\\
&=&\sigma[x,[y,a]_\g]_\g-\nu([x,[y,a]_\g)-\mu_0(x)\nu(y,a)+\sigma[y,[a,x]_\g]_\g-\nu(y,[a,x]_\g)-\mu_0(y)\nu(a,x)\\
&&+\sigma[a,[x,y]_\g]_\g-\nu(a,[x,y]_\g)-\mu_1(a)\omega(x,y)\\
&=&\sigma l_3^\g(x,y,\dM_\g
a)-\mu_0(x)\nu(y,a)-\mu_0(y)\nu(a,x)-\mu_1(a)\omega(x,y)\\
&&-\nu(x,[y,a]_\g)-\nu(y,[a,x]_\g)-\nu(a,[x,y]_\g)\\
&=&\hat{l_{3}}(\sigma x,\sigma y,\sigma(\dM_\g a)+\varphi(a))\\
&=&\sigma l_3^\g(x,y,\dM_\g a)-\theta(x,y,\dM_\g
a)+\mu_2(x,y)\varphi(a),
\end{eqnarray*}
which implies that
\begin{eqnarray}
\label{eq:con6}\nonumber&&\DERbr{f_0(x),f_2^1(y,a)}+\DERbr{f_0(y),f_2^1(a,x)}-\DERbr{f_1(a),f_2^0(x,y)}\\
\nonumber&=&\DERbr{\mu_0(x),\nu(y,a)}+\DERbr{\mu_0(y),\nu(a,x)}-\DERbr{(\mu_1-\varphi)(a),(-\mu_2+\omega)(x,y)}\\
\nonumber&=&\nu([x,y]_\g,a)+\nu([y,a]_\g,x)+\nu([a,x]_\g,y)+\theta(x,y,\dM_\g
a)\\
&=&f_2^1([x,y]_\g,a)+f_2^1([y,a]_\g,x)+f_2^1([a,x]_\g,y)+f_3(x,y,\dM_\g
a).
\end{eqnarray}

Since for any $x,y,z,t\in\g_0,$
$$
\hat{l_{3}}([\sigma x,\sigma y]_{\hat{\g}},\sigma z,\sigma
t)+c.p.=[\sigma x,\hat{l_{3}}(\sigma y,\sigma z,\sigma
t)]_{\hat{\mathfrak{g}}}+c.p..
$$
The left hand side  is equal to
\begin{eqnarray*}
&&\hat{l_3}(\sigma[x,y]_\g-\omega(x,y),\sigma z,\sigma t)+c.p.\\
&=&\sigma
l_3^\g([x,y]_\g,z,t)-\theta([x,y]_\g,z,t)-\mu_2(z,t)\omega(x,y)+c.p.,
\end{eqnarray*}
and the right hand side is equal to
\begin{eqnarray*}
&&[\sigma x,\sigma l_3^\g(y,z,t)-\theta(y,z,t)]_{\hat{\g}}+c.p.\\
&=&\sigma[x,l_3^\g(y,z,t)]_\g-\nu(x,l_3^\g(y,z,t))-\mu_0(x)\theta(y,z,t)+c.p..
\end{eqnarray*}
Thus, we have
\begin{eqnarray}
 \label{eq:con7}\nonumber&& \DERbr{f_0(x),f_3(y,z,t)}+f_2^1(x,l_3^\g(y,z,t))+c.p.\\&=&\nonumber\DERbr{\mu_0(x),\theta(y,z,t)}+\nu(x,l_3^\g(y,z,t))+c.p.\\
  \nonumber&=&\theta([x,y]_\g,z,t)+\mu_2(z,t)\omega(x,y)+c.p.\\
  &=&f_3([x,y]_\g,z,t)+\DERbr{f_2^0(x,y),f_2^0(z,t)}+c.p..
\end{eqnarray}

By \eqref{eq:con1},
\eqref{eq:con2},\eqref{eq:con3},\eqref{eq:con4},\eqref{eq:con5},\eqref{eq:con6},\eqref{eq:con7},
we obtain that $f$ is a morphism from $\frkg$ to $\DER(\frkh)$.

Given another splitting $\sigma'$ of the extension, there are the
induced linear maps
$(\varphi',\mu_{0}',\mu_{1}',\mu_{2}',\omega',\nu',\theta')$ such
that
$$f'=(f_0',f_1',{f_2^0}',{f_2^1}',f_3')=(\mu_{0}',\mu_{1}'-\varphi',-\mu_{2}'+\omega',\nu',\theta')$$
is a morphism from $\g$ to $\DER(\frkh)$. Assume that
$$
\sigma(x)=\sigma'(x)+b_0(x),\quad \sigma(a)=\sigma'(a)+b_1(a),
$$
where $b_0:\g_0\longrightarrow\frkh_0$ and
$b_1:\g_1\longrightarrow\frkh_1$  are linear maps. Then it is
straightforward to deduce that
\begin{eqnarray*}
\mu_0(x)-\mu_0'(x)&=&\ad(b_0(x)),\\
 \varphi'(a)-\varphi(a)&=&b_0(\dM_\g a)-\dM_\frkh b_1(a),\\
 \mu_1(a)-\mu_1'(a)&=&\ad_{b_1(a)},\\
 (\mu_2-\mu_2')(x,y)&=&l_{\mu_0'(x)}(b_0(y),\cdot)-l_{\mu_0'(y)}(b_0(x),\cdot)+l_3^\frkh(b_0(x),b_0(y),\cdot),\\
 (\omega'-\omega)(x,y)
&=&\mu_0'(x)b_0(y)-\mu_0'(y)b_0(x)+[b_0(x),b_0(y)]_\frkh-b_0[x,y]_\g+\dM_\frkh\circ
b_2(x,y),\\
(\nu'-\nu)(x,a)&=&\mu_{0}'(x)(b_{1}(a))-\mu_{1}'(a)(b_{0}(x))-b_1([x,a]_\g)+[b_{0}(x),b_{1}(a)]_\frkh,\\
 (\theta'-\theta)(x,y,z)&=&\mu_2'(x,y)b_0(z)+l_{\mu_0'(x)}(b_0(y),b_0(z))+c.p.-b_1(l_3^\g(x,y,z))+l_3^\frkh(b_0(x),b_0(y),b_0(z)).
\end{eqnarray*}
Then it is straightforward to see that $f'$ is equivalent to $f$ via
$(b_0,b_1,b_2=0)$. \qed\vspace{3mm}

Thus by choosing a splitting,  we can transfer the Lie 2-algebra
structure on $\hat{\g}$ to $\g\oplus\frkh$, which we denote by
 $(\widehat{\g\oplus\frkh},\hat{\dM},[\cdot,\cdot]^{\widehat{}},\hat{l_3})$:
\begin{equation}\label{eq:bracket}
\left\{\begin{array}{rcl}
\hat{\dM}(a+m)&\triangleq&\dM_\g(a)+\varphi(a)+\dM_\frkh(m),\\
\brh{x+u,y+v}&\triangleq&[x,y]_\g-\omega(x,y)+\mu_{0}(x)v-\mu_{0}(y)u+[u,v]_\frkh,\\
\brh{x+u,a+m}&\triangleq&[x,a]_\g-\nu(x,a)+\mu_{0}(x)m-\mu_{1}(a)u+[u,m]_\frkh,\\
 \hat{l_3}(x+u,y+v,z+w)&\triangleq& l_3^\g(x,y,z)-\theta(x,y,z)+l_3^\frkh(u,v,w)\\
&&+\mu_{2}(x,y)(w)+\mu_{2}(z,x)(v)+\mu_{2}(y,z)(u)\\
&&+l_{\mu_{0}(x)}(v,w)+l_{\mu_{0}(y)}(w,u)+l_{\mu_{0}(z)}(u,v),
\end{array}\right.
\end{equation}
for any $x,y,z\in \g_0$, $u,v,w\in\frkh_0$, $a\in\g_1$ and
$m\in\frkh_1$.

Thus any extension $E_{\hat{\g}}$ given by \eqref{eq:ext1} is
isomorphic to
\begin{equation}\label{eq:ext2}
\CD
  0 @>0>>  \frkh_1 @>i_1>> \g_1\oplus \frkh_1 @>p_1>> \g_1 @>0>> 0 \\
  @V 0 VV @V \dM_\frkh VV @V \hat{\dM} VV @V\dM_\g VV @V0VV  \\
  0 @>0>> \frkh_{0} @>i_0>> \g_0\oplus \frkh_0 @>p_0>> \g_0@>0>>0,
\endCD
\end{equation}
where the Lie 2-algebra structure on $\frkg\oplus\frkh$ is given by
\eqref{eq:bracket} for some morphism \eqref{eq:morphism},
$(i_0,i_1)$ is the inclusion and $(p_0,p_1)$ is the projection. We
denote the extension \eqref{eq:ext2} by
$\widehat{\E}_{\frkg\oplus\frkh}$.

\begin{thm} \label{mainthm}
There is a 1-1 correspondence between isomorphism classes of
extensions of Lie 2-algebras given by \eqref{eq:ext2} and
equivalence classes of morphisms $\eqref{eq:morphism} $ from the Lie
2-algebra $\frkg$ to the derivation Lie 3-algebra $\DER(\frkh)$.
\end{thm}

\pf Given two isomorphic extensions $\hat{\E}_{\g\oplus\frkh}$ and
${\E}'_{\g\oplus\frkh}$. Let
$F=(F_0,F_1,F_2):\hat{\E}_{\g\oplus\frkh}\longrightarrow{\E}'_{\g\oplus\frkh}$
be the corresponding isomorphism. By choosing two splittings
$\sigma$ and $\sigma'$ respectively, we get two morphisms $f$ and
$f'$ from $\g$ to $\DER(\frkh)$. In the following, we prove that
$f'$ is  equivalent to $f$.

 Since $F$ is an isomorphism of extensions, we have
$$F_2(u,v)=0,\quad F_2(x,u)=0,\quad F_2(x,y)\in\frkh_1,$$
and there exist two linear maps $\psi_0:\g_0\longrightarrow\frkh_0$
and $\psi_1:\g_1\longrightarrow\frkh_1$ such that
 $$F_0(x+u)=x+\psi_0(x)+u,\quad
F_1(a+m)=a+\psi_1(a)+m.$$ Set $b_0=\psi_0,b_1=\psi_1$ and $b_2=F_2$.


By
\begin{eqnarray*}\label{eqn:F1F0}
F_0(\brh{x,u})-\brt{F_0(x),F_0(u)}&=&0,\\
F_1(\brh{x,m})-\brt{F_0(x),F_1(m)}&=&0,
\end{eqnarray*}
we get
\begin{eqnarray*}
\label{eq:mumu}(\mu_0(x)-\mu_0'(x))(u)&=&\ad_{\psi_0(x)}(u),\\
\nonumber(\mu_0(x)-\mu_0'(x))(m)&=&\ad_{\psi_0(x)}(m),
\end{eqnarray*}
which implies that
\begin{equation}\label{eq:mainf0}
(f_0-f_0')(x)=\mu_0(x)-\mu_0'(x)=\dM_\D(\psi_0(x))=\dM_\D(b_0(x)).
\end{equation}

We also have
\begin{eqnarray*}
\label{eqn:fi} \dM'F_1(a)&=&F_0\hat{\dM}(a),\\
\label{eqn:F0F1} F_1(\brh{u,a})-\brt{F_0(u),F_1(a)}&=&0,
\end{eqnarray*}
which implies  that
\begin{eqnarray*}
\varphi'(a)-\varphi(a)&=&\psi_0(\dM_\g a)-\dM_\frkh \psi_1(a),\\
 \quad\mu_1(a)-\mu_1'(a)&=&\ad_{\psi_1(a)}.
\end{eqnarray*}
Therefore, we have
\begin{eqnarray}\nonumber
(f_1-f_1')(a)&=&(\mu_{1}-\varphi)(a)-(\mu_{1}'-\varphi')(a)=\dM_\D(\psi_{1}(a))+\psi_{0}(\dM_\g a)\\
\label{eq:mainf1}&=&\dM_\D(b_{1}(a))+b_{0}(\dM_\g a).
\end{eqnarray}

 Furthermore, we have
$$
F_0\brh{x,y}-\brt{F_0(x),F_0(y)}=\dM'F_2(x,y),
$$
which implies that
\begin{eqnarray*}\label{eq:econ3}
\omega'(x,y)-\omega(x,y)
=\mu_0'(x)b_0(y)-\mu_0'(y)b_0(x)+[b_0(x),b_0(y)]_\frkh-b_0[x,y]_\g+\dM_\frkh\circ
b_2(x,y).
\end{eqnarray*}

Since $F$ is a Lie 2-algebra morphism, we have the equality:
$$\brt{F_0(x),F_2(y,u)}+c.p.+l_3'(F_0(x),F_0(y),F_0(u))=F_2(\brh{x,y},u)+c.p.+F_1\hat{l_3}(x,y,u).$$
The left hand side is equal to
\begin{eqnarray}\label{e8}
\nonumber&&-\ad_{F_2(x,y)}(u)+\mu_2'(x,y)(u)+l_3^\frkh(\psi_0(x),\psi_0(y),u)+l_{\mu_0'(x)}(\psi_0(y),u)+l_{\mu_0'(y)}(u,\psi_0(x)),
\end{eqnarray}
and the right hand side is equal to $\mu_2(x,y)(u),$
 which implies that
$$
 \mu_2(x,y)-\mu_2'(x,y)=-\ad_{b_2(x,y)}+l_{\mu_0'(x)}(b_0(y),\cdot)-l_{\mu_0'(y)}(b_0(x),\cdot)+l_3^\frkh(b_0(x),b_0(y),\cdot).
$$
Thus, we have
\begin{eqnarray}
\nonumber ({f_2^0}'-f_2^0)(x,y)&=&\big((\mu_2-\mu_2')(x,y),(\omega'-\omega)(x,y)\big)\\
\nonumber&=&\DERbr{\mu_0'(x),b_0(y)}-\DERbr{\mu_0'(y),b_0(x)}-b_0([x,y]_\g)\\
\nonumber&&-\dM_\D({b_2(x,y)})+\DERbr{\dM_\D(b_0(x)),b_0(y)}\\
\nonumber&=&\DERbr{f_0'(x),b_0(y)}-\DERbr{f_0'(y),b_0(x)}-b_0([x,y]_\g)\\
\label{eq:mainf20}&&-\dM_\D({b_2(x,y)})+\DERbr{\dM_\D(b_0(x)),b_0(y)}.
\end{eqnarray}

Similarly, by $F_1\brh{x,a}-\brt{F_0(x),F_1(a)}=F_2(x,\hat{\dM}a)$,
we get
\begin{eqnarray}
\nu'(x,a)-\nu(x,a)&=&\DERbr{\mu_{0}'(x),b_{1}(a)}+\DERbr{\mu_{1}'(a)-\varphi'(a),b_{0}(x)}-b_1([x,a]_\g)\nonumber\\
\nonumber &&+b_2(x,\dM_\mathfrak{g}a)+\DERbr{\dM_\D(b_{0}(x)),b_{1}(a)}\\
 &=&\DERbr{f_{0}'(x),b_{1}(a)}+\DERbr{f_{1}'(a),b_{0}(x)}-b_1([x,a]_\g)\nonumber\\
 \label{eq:mainnu}&&+b_2(x,\dM_\mathfrak{g}a)+\DERbr{\dM_\D(b_{0}(x)),b_{1}(a)}.
\end{eqnarray}

At last, by the equality
$$\brt{F_0(x),F_2(y,z)}+c.p.+l_3'(F_0(x),F_0(y),F_0(z))=F_2(\brh{x,y},z)+c.p.+F_1\hat{l_3}(x,y,z),$$
 we have
\begin{eqnarray}
\nonumber(\theta'-\theta)(x,y,z)&=&\mu_0'(x)b_2(y,z)-b_2([x,y]_\g,z)+c.p.\\
\nonumber&&+\mu_2'(x,y)b_0(z)+[b_0(x),b_2(y,z)]_\frkh+l_{\mu_0'(x)}(b_0(y),b_0(z))+c.p.\\
\label{eq:mainf3}&&-b_1(l_3^\g(x,y,z))+l_3^\frkh(b_0(x),b_0(y),b_0(z)).
\end{eqnarray}
 By \eqref{eq:mainf0},\eqref{eq:mainf1},\eqref{eq:mainf20},\eqref{eq:mainnu},\eqref{eq:mainf3}, we deduce that $f'$ and $f$ are
 equivalent.

Conversely, assume that
$f'=(\mu_{0}',\mu_{1}'-\varphi',-\mu_{2}'+\omega',\nu',\theta')$ is
equivalent to
$f=(\mu_{0},\mu_{1}-\varphi,-\mu_{2}+\omega,\nu,\theta)$ in the
sense of Definition \ref{defi:equivalent}. For any $u,v\in\frkh_0,
x,y\in\g_0, m\in\frkh_1$ and $a\in\g_1$, set
\begin{eqnarray*}
F_0(x+u)&=&x+b_0(x)+u,\\
F_1(a+m)&=&a+b_1(a)+m,\\
F_2(x+u,y+v)&=&b_2(x,y).
\end{eqnarray*}
By similar computations to the first part of the proof, we can
deduce that $F=(F_0,F_1,F_2)$ is an isomorphism from the extension
$\hat{\E}_{\g\oplus\frkh}$ to $\E'_{\g\oplus\frkh}$. This completes
the proof.
 \qed


\begin{thebibliography}{20}

\bibitem{AMR}
Alekseevsky D,  Michor P W,  Ruppert W.  Extensions of Lie algebras.
arXiv:math.DG/0005042.

\bibitem{AMR2}
 Alekseevsky D,  Michor P W,  Ruppert W. Extensions of Super Lie algebras. \emph{J. Lie Theory} 15(1):125-134   (2005).

\bibitem{Baez}
 Baez J C,  Crans A S.  Higher-dimensional algebra. VI. Lie 2-algebras. \emph{Theory Appl.Categ.}, 12:492-538 (electronic) (2004).

 \bibitem{brahic}
 Brahic O. Extensions of Lie brackets. \emph{J. Geom. Phys.} 60, no. 2, 352-374 (2010).

\bibitem{Malte}
Dehling M. Shifted $L_\infty$-bialgebras. master thesis, G\"ottingen
University, 2011.

\bibitem{Eilenberg}
  Eilenberg S,  Maclane S.:Cohomology theory in abstract groups. II. Group extensions with non-abelian kernel. \emph{Ann. Math.}, 48:326-341 (1947).


 \bibitem{Hochschild}Hochschild G. Cohomology clases of finite type and finite dimensional kernels for Lie
 algebras.
\emph{Am. J. Math.} 76, 763-778 (1954).

\bibitem{Nonabeliancohomology}
Inassaridze N,  Khmaladze E,  Ladra M. Non-abelian cohomology and
extensions of Lie algebras. \emph{Journal of Lie Theory}, 18:413-432
(2008).

\bibitem{Lada1}
 Lada T,  Stasheff J. Introduction to sh Lie algebras for
physicists. \emph{Int. J.
              Theo. Phys.}, Vol. 32(7):1087-1103 (1993).

\bibitem{Lada2}
 Lada T, Markl M.  Strongly homotopy Lie algebras.  \emph{Comm.
Alg.}, 23(6):2147-2161 (1995).

\bibitem{Mackenzie}
 Mackenzie K. \emph{ Lie groupoids and Lie algebroids in diferential
geometry}. London Mathematical Society Lecture Note Series, 124,
Cambridge University Press, 1987.

\bibitem{stasheff2}   Schreiber U, Stasheff J. Structure of Lie n-Algebras. unpublished work.


\bibitem{sheng1}
Sheng Y, Zhu C. Semidirect products of representations up to
homotopy. \emph{Pacific J.
  Math.}, 249 (1), 211-236 (2011).


\bibitem{sheng2}
  Sheng Y,  Liu Z-J,  Zhu C. Omni-Lie 2-algebras and their Dirac structures.  \emph{J. Geom. Phys.}, 61:560-575,
 (2011).

\bibitem{sheng3}
  Sheng Y,  Zhu C. Higher Extensions of Lie Algebroids and Application to Courant Algebroids. arXiv:1103.5920v1.

\bibitem{Shukla} Shukla  U. A cohomology for Lie algebras. \emph{J. Math. Soc.
Japan} 18, 275-289 (1966).

\bibitem{stevenson}
Stevenson D. Schreier Theory for Lie 2-algebras. unpublished work.




\end{thebibliography}
\end{document}